\documentclass[epj]{svjour}
\usepackage{graphicx}
\usepackage{amsmath}
\usepackage{ifthen}
\usepackage{amssymb}
\usepackage{color}
\newcommand{\slaninacolor}{true}
\def\slaninafigdir{.}
\ifthenelse{\equal{\slaninacolor}{true}}{%
\def\slaninalineone{{\textcolor{red}{red line}}}
\def\slaninalinetwo{{\textcolor{blue}{blue line}}}
\def\slaninalinethree{{\textcolor{green}{green line}}}
\def\slaninalinefour{{\textcolor{magenta}{magenta line}}}
\def\slaninalineonethree{{\textcolor{green}{green line}}}
\def\slaninalinethreefour{{\textcolor{green}{green line}}}
\def\slaninacolorinname{-color}

}{
\renewcommand{\textcolor}[2]{{#2}}
\renewcommand{\color}[1]{{}}
\def\slaninalineone{{solid line}}
\def\slaninalinetwo{{dashed line}}
\def\slaninalinethree{{dash-dotted line}}
\def\slaninalinefour{{dotted line}}
\def\slaninalineonethree{{solid line}}
\def\slaninalinethreefour{{dotted line}}
\def\slaninacolorinname{}
}
\begin{document}
\title{%
Glass transition in a simple stochasic model with back-reaction 
}
\author{%
Franti\v{s}ek Slanina%
    \inst{1}\thanks{e-mail: {\tt slanina@fzu.cz}}
\and
Petr Chvosta%
    \inst{2}\thanks{e-mail: {\tt chvosta@kmf.troja.mff.cuni.cz}}
}
\institute{%
        Institute of Physics,
	Academy of Sciences of the Czech Republic,\\
	Na~Slovance~2, CZ-18221~Praha,
	Czech Republic
\and%
Department of Macromolecular Physics,
Faculty of Mathematics and Physics,
Charles University,\\
 V Hole\v{s}ovi\v{c}k\'ach~2, CZ-180~00~Praha,
Czech Republic
}%
\date{%
}
%
\abstract{%
We formulate and solve a model of dynamical arrest in colloids.
A particle is coupled to the bath of 
statistically identical particles. The dynamics is described by
Langevin equation with stochastic external force described by 
telegraphic noise. The interaction with the bath is taken into account
self-consistently through the back-reaction mechanism. Dynamically induced 
glass
transition occurs 
for certain value of the coupling strength. 
Edwards-Anderson parameter jumps discontinuously at
the transition. Another order parameter can be also defined, which
vanishes continuously with exponent $1/2$ at the critical point.
Non-linear response to harmonic perturbation is found.
} 
\maketitle
\section{Introduction} 

Glass transition and slow relaxation in systems characterised by weak
ergodicity breaking remains  still an open area, despite many efforts
and numerous significant results in the last decade 
\cite{angell_95,pa_ste_ab_an_84,pa_me_dedo_85,ki_thi_87,ki_thi_87a,ki_thi_wo_89,ritort_95,fra_he_95,cu_ku_93,fra_me_pa_pe_98,ba_bu_me_96a,kurchan_92,cu_ku_95,ma_pa_ri_94b,nieuwenhuizen_98,ca_gia_pa_99,ca_gia_pa_99a,me_pa_99,co_pa_ve_00,cu_ku_pa_ri_95,par_sla_00,kaw_kim_01}.
Among the host of diverse phenomena we are motivated here mostly by
the effect of dynamical arrest in colloidal matter 
\cite{fie_sol_cat_99,fu_cat_03,cates_02,pue_fu_cat_02,fu_cat_02,zac_fof_daw_bul_sci_tar_02,law_rea_mcc_deg_tar_daw_02,sti_vla_lop_roo_mei_02,fab_got_sci_tar_thi_99},
observed experimentally and thoroughly investigated by numerical
simulations and Mode-Coupling method. Below the transition point, the
dynamics effectively leads to a glassy state with diverging viscosity,
however the static thermodynamic transition may not be identifiable. 
Indeed, dynamical or structural
arrest demonstrates the glass transition as a purely dynamic and
self-consistent 
phenomenon, where casual slowdown of certain particles prevents some other
particles from moving, which may slow down the others even more
etc. The self-consistent nature of the phenomenon is reflected by  
the analytical approaches available now. One of the most striking
phenomena in colloids, suspensions and granular matter is the
non-Newtonian response to mechanical perturbation. On one hand, we can
have
shear-thinning, which amounts to a decrease if viscosity due to
applied field, which can be interpreted as restoration of ergodicity
due to perturbation \cite{cates_02}. On the other hand, increase of
viscosity may result in shear thickening or even jamming, typically
observed in particulate or granular matter
\cite{cat_wit_bou_cla_99,ber_bib_sch_02}.

From the theoretical side, the Mode-Coupling (MC) equations provide us
with a well-established  
framework, capable of explaining a good deal of experimental data
\cite{gotze_91,go_sj_92,bouchaud,sjogren_03}. The attempts to derive the MC equations starting
from the Hamiltonian of the system were successful in the
mean-field approximation. It was perhaps the $p$-spin spherical model
\cite{ki_thi_87,ki_thi_87a,ki_thi_wo_89,cu_ku_95}, 
where the machinery reached the farther edge of our current
understanding of the phenomenon.
 
However, the bottom-up approach starting with writing explicit
Hamiltonians is far from being complete. The presence of the
reparametrisation invariance \cite{cu_ku_95,cu_ku_98} leaves the numerical solution
of the MC equations as the only means for obtaining the true
time-dependence of the correlation and response functions. Also the
mean-field approximation generally used now seems to be very difficult
to overcome.

The serious difficulties remaining in using the more advanced MC
techniques leave the space for more simple phenomenological
approaches. We want to follow this path in the present work.
 
Indeed, the mathematical substance of the Mode Coupling method can be
summarised by saying that the time dependence of the correlation (and response)
functions  depends non-linearly and in time-delayed manner on these
functions themselves. 
Actually, the memory kernel in the MC equations, which is primarily
dictated by the properties of the reservoir, depends of the system
dynamics.

We 
may represent the dynamics of the system by a stochastic process 
and the parameters of the process depend on time through the 
averaged
properties of the process itself. In order to study generic properties
of  such problems it can be useful to establish a simple
idealised model, which would capture the essential mathematic
ingredients while avoiding the complications which arise from choosing
a specific Hamiltonian at the beginning. The most important ingredient
in such an idealised model should be the mechanism of the 
{\it back-reaction}.

We introduced recently \cite{chv_sla_02} a very simple stochastic
process, in which the back-reaction leads to rich dynamic
behaviour. The main characteristics was the presence of a phase
transition from ergodic to non-ergodic phase. 
The principal aim of the present work is to
investigate analytically some of the properties of the transition and
from the numerical solution of the corresponding differential
equations infer the non-trivial critical behaviour. 

\section{Langevin equation with back--reaction}

\subsection{System of coupled particles}

The model system we will have in mind will be composed of particles,
relaxing to their equilibrium positions under the influence of
surrounding particles. They can be viewed as colloidal particles
immersed in a solvent, but the formulation of our model is generic
enough to allow for other interpretations as well, e. g. they can be
viewed as micro-domains in a relaxor ferroelectric material.

The time evolution of the model can lead to dynamical arrest, where
particles are locked in their positions by surrounding particles,
which are also locked in their turn. Therefore, the dynamics can lead
to the 
spatially disordered but time-stable stationary state with glass
properties. The indication of the glass transition will be 
the non-zero value of
the Edwards-Anderson order parameter and sensitivity to initial
conditions.
The interaction between particles will be taken into account on a
phenomenological level; if we concentrate on a randomly selected
particle (single relaxor), the external field from the rest of the
system (reservoir)  will change as the states of the other particles
(relaxors in reservoir) 
evolve. The changes in the local external field will be the more rapid the
faster is the evolution of the other particles. This leads to the idea
of expressing the intensity of the changes in th external field through
the velocity of movement of the relaxors in the reservoir. As we
suppose all particles to be statistically identical, the movement of our
single relaxor should be in probabilistic sense equivalent to the
movement of any relaxor within the reservoir. This consideration
closes the loop. 

We will try to express the intensity of the changes 
of the local field through the averaged properties of the
movement of the single relaxor itself. This introduces the idea of
{\it back-reaction}: the probabilistic properties of 
the reservoir dictate the system evolution and the averaged system
dynamics tunes the properties of the reservoir itself.

To be more specific, our single relaxor will be described by the
continuous real stochastic variable ${\sf X}(t)$.  It will evolve
under influence of the environmental force, represented by the stochastic
variable ${\sf Q}(t)$. The force will be modelled by 
 a two--valued random process ${\sf Q}(t)\in\{-q,+q\}$  \cite{horsthemke},
jumping at random instants. Occurrence of the jumps are governed by
a self-exciting point process \cite{cox,snyder} with time-dependent
intensity $\frac{1}{2}\lambda(t)$. 

For given (friction-reduced) force ${\sf Q}(t)$ the single relaxor is
described by the Langevin equation \cite{pomeau,sibani,morita}
\begin{equation}
\frac{d}{dt}\,{\sf X}(t)=-\gamma\,{\sf X}(t)+{\sf Q}(t)
\label{eq:langevin}
\end{equation}
with initial conditions ${\sf X}(0) = {\sf X}_0$ and ${\sf Q}(0)={\sf
Q}_0$. 

We may consider the process ${\sf X}(t)$
as a movement of an over-damped particle which slides in
a parabolic potential well; the parabola jumps between 
two positions at random instants, with time-dependent rate
$\frac{1}{2}\lambda(t)$.

The intensity of the process, or the frequency of jumps,
$\frac{1}{2}\lambda(t)$, is related to
the movement of the surrounding relaxors, considered as a
reservoir. Intuitively the frequency 
must be smaller if the movement of the relaxors is slower. 
Therefore, the function $\lambda(t)$ should be coupled to the velocity
${\sf V}(t)=\frac{d}{dt}\,{\sf X}(t)$.

\begin{figure}[ht]
\centerline{
\includegraphics[scale=0.45]{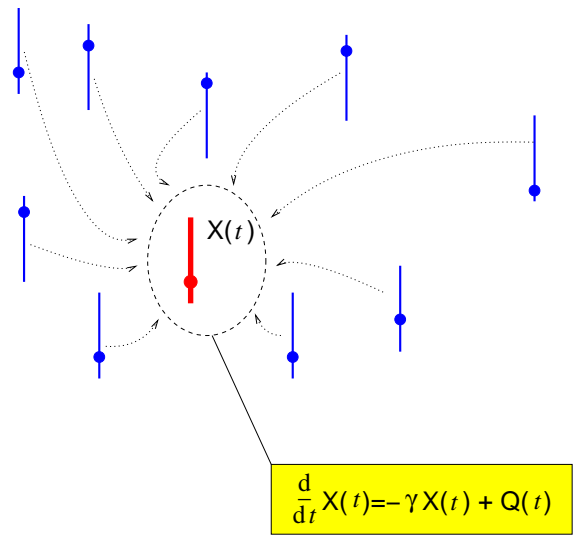}
}
\caption{Schematical picture of our model. Relaxing particles in the
reservoir influence the relaxation of the selected particle, described
by the Langevin equation shown in the frame box.}
\label{fig:backr-scheme}
\end{figure}

However, it is
not obvious a priori what should be the specific functional dependence.
We only require that the dependence is described by a non-negative
function analytic 
at the origin. The simplest choice satisfying this property is 
\begin{equation}
\lambda(t)=\epsilon\frac{\gamma}{q^{2}}\left\langle\,
{\sf V}^{2}(t)\,\right\rangle
\label{eq:lambda}
\end{equation}
where $\epsilon$ is the dimensionless coupling strength parameter.
The latter prescription is the form of the back-reaction we will study
in the following. 
The model described above is sketched schematically in the
Fig. \ref{fig:backr-scheme}.

The parameters $\gamma$ and $q$ can be in principle rescaled to 1 by
appropriate choice of the units of time and length. Therefore, the
coupling strength $\epsilon$ remains to be the only physically
relevant parameter tuning the behaviour of the system. As we will see,
there is a qualitative change in the behaviour of the system at a certain
critical value of $\epsilon$.

\subsection{Properties of the environmental force ${\sf Q}(t)$}

The force ${\sf Q}(t)$ is a time-inhomogeneous Markov process,
therefore its properties are fully described by a master equation.
More specifically, let us define the probabilities
\begin{equation}
\pi_{\pm}(t)={\rm Prob}\left\{\,{\sf Q}(t)=\pm q\right\}
\end{equation}
making the vector 
$\boldsymbol{\pi}(t)=\begin{pmatrix}\pi_-(t)\\\pi_+(t)\end{pmatrix}$
which satisfies the Pauli master equation
\begin{equation}
\frac{d}{dt}
\boldsymbol{\pi}(t)=
-\frac{1}{2}\lambda(t)
\left(
\begin{array}{rr}
1&-1\\-1&1
\end{array}
\right)
\boldsymbol{\pi}(t)\; .
\end{equation}
Solving the equation amounts to calculation of the corresponding
time-ordered exponential. The averages and correlation functions can
be expressed through the integrated intensity
\begin{equation}
\Lambda(t) = \int_0^t\,\lambda(t'){\rm d}t'\; .
\end{equation}
The time dependence of the vector $\boldsymbol{\pi}(t)$ can be written
through the 
semi-group operator 
\begin{equation}
R(t,t_0)=
\left(
\begin{array}{rr}
1&0\\0&1
\end{array}
\right)
+
\frac{1}{2}\left({\rm e}^{-\Lambda(t)+\Lambda(t_0)}-1\right)
\left(
\begin{array}{rr}
1&-1\\-1&1
\end{array}
\right)
\end{equation}
as $\boldsymbol{\pi}(t)=R(t,t_0)\boldsymbol{\pi}(t_0)$.
Note that the semi-group operator obeys
$
R(t,\tau)
R(\tau,t_0)=
R(t,t_0)\;\forall\tau\in[t_0,t]
$
which testifies the Markov property of the process.

More explicitly, we find
\begin{equation}
\langle {\sf Q}(t)\rangle=\langle {\sf Q}_0\rangle \exp(-\Lambda(t))
\label{eq:averageQ}
\end{equation}
\begin{equation}
\langle {\sf Q}(t){\sf Q}(t_1)\rangle=q^2 \exp(-|\Lambda(t)-\Lambda(t_1)|)
\label{eq:correlationQQ}
\end{equation}
Similarly, also the higher correlation functions can be written as
products of exponentials with combinations of $\Lambda(t)$ with
appropriate time arguments in the exponents. 
In fact, higher order correlation functions factorise into product of
first and second order correlations, e. g.
\begin{equation}
\langle \prod_{l=1}^{2k}{\sf Q}(t_l)\rangle
=
\prod_{l=1}^k\langle {\sf Q}(t_{2l}){\sf Q}(t_{2l-1})\rangle
\end{equation}
for 
$t_{2k}\ge t_{2k-1}\ge...\ge t_2\ge t_1$.
Another consequence is, that the cumulants of higher order than two vanish.

Note that the function $\Lambda(t)$ is non-decreasing and can either
diverge 
(if $\lim_{t\to\infty}\lambda(t)>0$) or assume a finite limit for
$t\to\infty$, if $\lambda(t)$ approaches 0 fast enough.

\section{Glass transition and asymptotic relaxation}
\label{sec:glasstransition}

\subsection{Equations for moments}
For any given realisation of the process ${\sf Q}(t)$,
the formal solution of
Eq. (\ref{eq:langevin}) is 
\begin{equation}
{\sf X}(t)={\sf X}_0\,{\rm e}^{-\gamma t}+\int_{0}^{t}\, {\rm
e}^{-\gamma(t-t')}{\sf Q}(t')\,{\rm d}t'\quad .
\label{eq:solutionforX}
\end{equation}

If the function $\lambda(t)$ were known, various moments (and
correlation functions) of the random process ${\sf X}(t)$ could have
been computed from (\ref{eq:solutionforX}) using the expressions
(\ref{eq:averageQ}) and 
(\ref{eq:correlationQQ}). However, in our case the function
$\lambda(t)$  should be computed from the condition (\ref{eq:lambda}),
relating it to the second moment of the time derivative of 
${\sf X}(t)$. This suggests that sufficiently broad set of moments of
${\sf X}(t)$ and ${\sf V}(t)$  may provide a closed set of ordinary
differential equations. The solution of this set will yield the
closed description of the behaviour of our model.

Indeed, we can define four auxiliary functions 
\begin{eqnarray}
\label{sdef1}
s_{1}(t)&=&{\rm e}^{-\Lambda(t)}\\
\label{sdef2}
s_{2}(t)&=&{\rm e}^{-\gamma t}\int_{0}^{t}{\rm d}t'
\,{\rm e}^{\gamma t'-\Lambda(t')}\,\\
\label{sdef3}
s_{3}(t)&=&{\rm e}^{-\gamma t-\Lambda(t)}
\int_{0}^{t}{\rm d}t'\,{\rm e}^{\gamma t'+\Lambda(t')}\,\\
\label{sdef4}
s_{4}(t)&=&{\rm e}^{-2\gamma t}\int_{0}^{t}{\rm d}t'\, {\rm e}^{\gamma
t'-\Lambda(t')}\,   
\int_{0}^{t'}{\rm d}t''\,{\rm e}^{\gamma t''+\Lambda(t'')}
\end{eqnarray}
and express the requested quantities through these functions. For
example the average coordinate can be written as
\begin{equation}
\langle{\sf X}(t)\rangle = \langle{\sf X}_0\rangle\, 
{\rm e}^{-\gamma t}
+\langle{\sf Q}_0\rangle\,s_2(t) \; .
\label{eq:averageX}
\end{equation}
 Similarly, the second moment of
the coordinate is
\begin{equation}
\langle{\sf X}^2(t)\rangle = \langle{\sf X}^2_0\rangle\, 
{\rm e}^{-2\gamma t}
+2\langle{\sf X}_0{\sf Q}_0\rangle\,{\rm e}^{-\gamma t}\,s_2(t) 
+2q^2\,s_4(t)\, .
\end{equation}
The functions $s_1(t)$ to $s_4(t)$ can be found by solving the set of
non-linear differential equations 
\begin{eqnarray}
\label{seqn1}
\dot{s}_{1}(t)&=&-\lambda(t)\,s_{1}(t)\\
\label{seqn2}
\dot{s}_{2}(t)&=&-\gamma\, s_{2}(t)+s_{1}(t)\\
\label{seqn3}
\dot{s}_{3}(t)&=&1-\left(\gamma+\lambda(t)\right)\,s_{3}(t)\\
\label{seqn4}
\dot{s}_{4}(t)&=&-2\gamma\, s_{4}(t)+s_{3}(t)
\end{eqnarray}
with initial conditions $s_1(0)=1$, $s_2(0)=s_3(0)=s_4(0)=0$.
The function $\lambda(t)$ occurring in the latter equations is itself
a combination of the functions $s_1(t)$ to $s_4(t)$ 
\begin{equation}
\begin{split}
\lambda(t)&=\epsilon
\left[\gamma-2\gamma^2\,s_3(t)+2\gamma^3\,s_4(t)\right]+\\
&+\frac{\epsilon\gamma^2}{q^2}\left[\gamma\langle{\sf X}^2_0\rangle\,
{\rm e}^{-2\gamma t} + 2\gamma\langle{\sf X}_0{\sf Q}_0\rangle\,
{\rm e}^{-\gamma t}\, s_2(t)-\right.\\
&\left.-2\langle{\sf X}_0{\sf Q}_0\rangle{\rm
e}^{-\gamma t}\, s_1(t)\right] \quad . 
\end{split}
\label{eq:lambdathroughsses}
\end{equation}

In the following we will assume that the initial condition of the
stochastic process is ${\sf X}_0=0$ and $\langle{\sf
Q}_0\rangle=0$, except explicitly mentioned cases. 
This lead to significant 
simplification of the mathematical structure of the equations. Indeed,
the equations for $s_3(t)$ and $s_4(t)$ form a closed pair of equations
\begin{equation}
\begin{split}
\dot{s}_3(t)=&1-(1+\epsilon)\gamma\,s_3(t)+\\
             &+2\epsilon\gamma^2\,(s_3(t)-\gamma\,s_4(t))\,s_3(t)    \\
\dot{s}_4(t)=&s_3(t)-2\gamma\,s_4(t)\quad .
\end{split}
\label{eq:fors3s4}
\end{equation}
Unfortunately, the system (\ref{eq:fors3s4}) cannot be solved
analytically. The best one can do is to transform the non-linear
Ricatti-type set (\ref{eq:fors3s4}) to one differential
equation of Abel type, 
whose solution, however, is not generally known. Therefore, we will
solve the equations (\ref{eq:fors3s4}) numerically. Nevertheless,
there is still a significant 
amount  of information which can be extracted analytically.

\begin{figure}[t]
\includegraphics[scale=0.8]{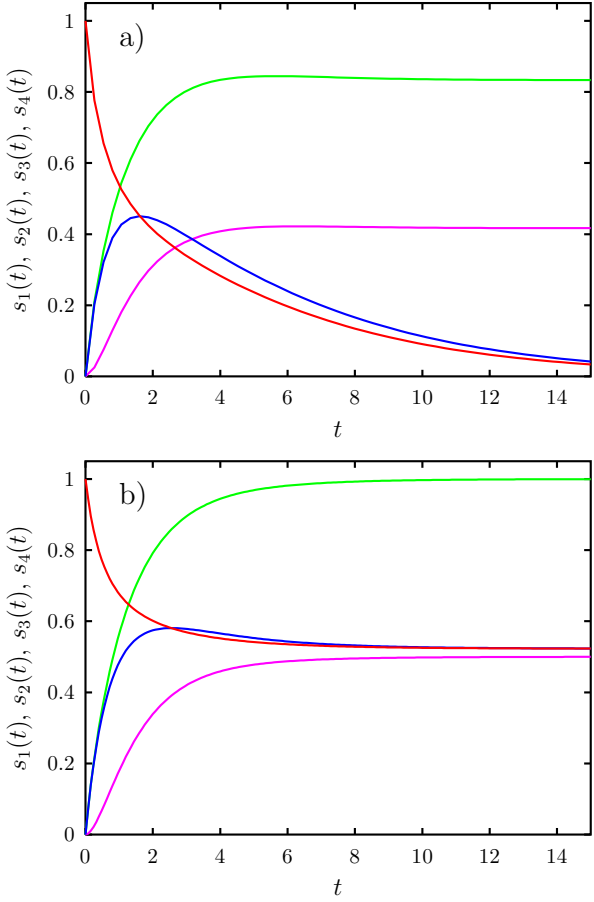}
\caption{Time evolution of the auxiliary functions $s_1(t)$ 
(\slaninalineone), $s_2(t)$ (\slaninalinetwo), $s_3(t)$
(\slaninalinethree), and   
$s_4(t)$ (\slaninalinefour) for $\gamma-1$, $q=1$. The panel a) 
corresponds to the value of the parameter $\epsilon=1.2$, while in the
panel b) we have  $\epsilon=0.8$.}
\label{fig:s1-s4}
\end{figure}

\begin{figure}[t]
\includegraphics[scale=0.8]{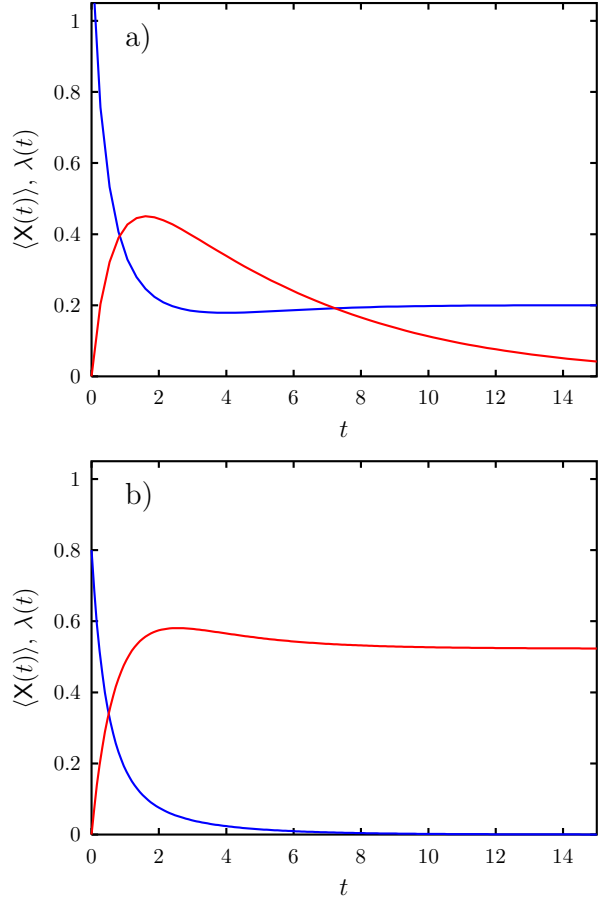}
\caption{Time evolution of the 
average coordinate $\langle{\sf X}(t)\rangle$ 
computed for initial condition $\langle{\sf Q}_0\rangle=1$ 
(\slaninalineone) 
and of the the switching rate $\lambda(t)$ 
(\slaninalinetwo)
 for $\gamma=1$, $q=1$. The panel a)
corresponds to the value of the parameter $\epsilon=1.2$, while in the
panel b) we have  $\epsilon=0.8$.}
\label{fig:evolution}
\end{figure}

The typical results of numerical solution are shown in 
figures \ref{fig:s1-s4} and \ref{fig:evolution}. In
Fig. \ref{fig:s1-s4} we can see the time evolution of the auxiliary
functions $s_1(t)$ to $s_4(t)$. Fig. \ref{fig:evolution} shows the
evolution of the switching rate $\lambda(t)$ and average coordinate
$\langle{\sf X}(t)\rangle$ for non-zero value of the initial
condition $\langle{\sf Q}_0\rangle$. We can observe qualitatively
different behaviour for $\epsilon<1$ and $\epsilon>1$: first, the
switching rate approaches a non-zero limit for $\epsilon>1$, while for
$\epsilon<1$ it decays to zero. This means that in the latter case the
system effectively freezes. This is further confirmed by the
observation that for $\epsilon<1$ the limit value of the average
coordinate depends on the initial conditions, while in the opposite
case the dependence on initial conditions is lost for large times, the
system equilibrates and
the average coordinate converges always to zero. The following
sections are mainly devoted to the analytical investigation of the above
observations.  

\subsection{Fixed points}
\label{sec:fixedpoits}
The first step in investigating the behaviour of the system
(\ref{eq:fors3s4}) is the search for the fixed points $[s_3^*,s_4^*]$ 
of the
dynamics.  
We found that there are only two fixed points, namely
\begin{equation}
[s_3^*,s_4^*]
=
[\frac{1}{\gamma\epsilon},\frac{1}{2\gamma^2\epsilon}]
\label{eq:fixedpointergodic}
\end{equation}
and
\begin{equation}
[s_3^*,s_4^*]
=
[\frac{1}{\gamma},\frac{1}{2\gamma^2}]\quad .
\label{eq:fixedpointnonergodic}
\end{equation}
Let us denote $\lambda_\infty$ the value of $\lambda(t)$
calculated at the corresponding fixed point. Using
(\ref{eq:lambdathroughsses}) the fixed point
(\ref{eq:fixedpointergodic}) yields 
$\lambda_\infty=(\epsilon-1)\gamma$ and the fixed point
(\ref{eq:fixedpointnonergodic}) yields $\lambda_\infty=0$.

The linear stability analysis reveals that for $\epsilon>1$ the fixed
point (\ref{eq:fixedpointergodic}) is stable, while
(\ref{eq:fixedpointnonergodic}) is unstable. On the other hand, for
$\epsilon<1$ the fixed 
point (\ref{eq:fixedpointnonergodic}) is stable, while
(\ref{eq:fixedpointergodic}) is unstable.
The case $\epsilon=1$ is a marginal one, where both fixed points 
have one of the eigenvalues equal to 0.
Therefore, the value $\epsilon=1$ marks a transition, whose nature will
be further pursued in the following.

\subsection{Ergodic regime $\epsilon> 1$}

In this case the relevant fixed point is (\ref{eq:fixedpointergodic}) 
and inserting its value to the expressions for the moments of 
${\sf X}(t)$ we find that both the average coordinate and the average velocity
relaxes to zero. On the other hand, the fluctuations of the coordinate
reach positive value, so
\begin{equation}
\begin{split}
\lim_{t\to\infty}\langle{\sf X}(t)\rangle&=0
\\
\lim_{t\to\infty}\langle{\sf X}^2(t)\rangle&=\frac{q^2}{\gamma^2\epsilon}
\quad .
\end{split}
\end{equation}
The external force switching rate converges to positive constant
$\lim_{t\to\infty}\lambda(t)=\gamma\,(\epsilon-1)$.
In all cases the quantities of interest converge exponentially to
their limit values
The rate of convergence is determined by the lowest in absolute value
eigenvalue, which is
\begin{equation}
\mu_1=-\frac{\gamma}{2}\left(\epsilon-\sqrt{\epsilon^2-8\epsilon+8}\right)\; .
\label{eq:mu1ergodic}
\end{equation}
In the interval $\epsilon\in(4-2\sqrt{2},4+2\sqrt{2})$ the eigenvalue
acquires a non-zero imaginary part, which means that oscillatory
behaviour is superimposed over the exponential relaxation.

The overall picture is the following. The back-reaction leads to
self-adjustment of the switching rate of the external force exerted by the 
reservoir. The coordinate fluctuates around the origin and these
fluctuations are stationary. Therefore, the stationary regime of the
system corresponds to the primitive version with fixed $\lambda$,
except the fact that the value of $\lambda$ is not given from outside,
but tuned by the dynamics itself. We call this regime ergodic, because
the particles do not freeze at some value of the coordinate ${\sf X(t)}$
but fluctuate forever.

The probability density for the coordinate
\begin{equation}
P(x,t)=\frac{\rm d}{{\rm
d} x}{\rm Prob}\{{\sf X}(t)\le x\}
\end{equation}
approaches for $t\to\infty$ the function 
\cite{chv_sla_02,morita}
\begin{equation}
\label{density}
\lim_{t\rightarrow\infty}P(x,t)=
\frac{\gamma}{q}\,\frac{(1-\widetilde{x}^{2})^
{(\epsilon-3)/2}}{B\left(\frac{\epsilon-1}{2},\frac{1}{2}\right)}\,\,
\Theta(1-\widetilde{x}^{2})\,\,,
\end{equation}
where $\widetilde{x}=x\gamma/q$, $\Theta(a)$ is the Heaviside
unit-step function, and 
$B(a,b)$ denotes the Beta-function \cite{abramowitz}.
We can observe a qualitative change at the value $\epsilon=3$. For
$\epsilon>3$ the limiting distribution (\ref{density}) has a maximum for
$x=0$ and approaches $0$ at the edges of the support
$[-q/\gamma,q/\gamma]$, while for $\epsilon<3$ it has a minimum at
$x=0$ and diverges at the edges of the support. The tendency for
accumulating the probability close to the points $\pm q/\gamma$ when
$\epsilon$ decreases can be regarded as a precursory phenomenon of the
transition to the non-ergodic regime, investigated in the next
sub-section. 

\subsection{Non-ergodic regime $\epsilon<1$}
\label{sec:nonergodic}

In this case we have (\ref{eq:fixedpointnonergodic}) as stable fixed
point. For $\langle{\sf Q}_0\rangle=0$ the average coordinate converges
to 0 again, but the   
second moment approaches the $\epsilon$-independent maximum value
\begin{equation}
\lim_{t\to\infty}\langle{\sf X}^2(t)\rangle=\frac{q^2}{\gamma^2}\; .
\label{eq:X2nonergodic}
\end{equation}
As the probability density for the coordinate
$P(x,t)$
 has support limited to the interval
$[-q/\gamma,q/\gamma]$, it follows from (\ref{eq:X2nonergodic})  the
the limiting 
probability density is composed of two $\delta$-functions of equal
weight $\frac{1}{2}$ located at
the edges of the latter interval.
More generally, for non-symmetric initial condition for the noise,
$\langle{\sf Q}_0\rangle\ne 0$, the
limiting probability density is the sum of two
$\delta$-functions, 
\begin{equation}
\lim_{t\to\infty}P(x,t)=\rho_+\,\delta\left(x-\frac{q}{\gamma}\right)
+\rho_-\,
\delta\left(x+\frac{q}{\gamma}\right)
\end{equation}
 the weights of which depend non-trivially on
 $\epsilon$ and the initial condition
\begin{equation}
\rho_\pm=\frac{1}{2}\left(1\pm\frac{\langle{\sf Q}_0\rangle}{q}\sigma(\epsilon)\right)
\end{equation}
where $\sigma(\epsilon)=\lim_{t\to\infty}s_1(t)$.

The function $\lambda(t)$ relaxes to zero. 
Therefore, in this regime, the switching of the external force
asymptotically stops and the coordinate ${\sf X}(t)$ approaches either
the
value  $+q/\gamma$ or $-q/\gamma$, where it freezes. So, the
coordinate acquires a random but time-independent asymptotic
value. More precisely, the mean coordinate $\langle{\sf X}(t)\rangle$
approaches a generally non-zero asymptotic value, which depends on the
initial condition.
This is the manifestation of glassy state in the regime 
$\epsilon<0$, characterised by broken ergodicity and non-zero
Edwards-Anderson order parameter. This point will be discussed more in
detail later in the presentation of correlation functions.

As in the ergodic phase, all quantities relax toward their limit values 
exponentially for large times. The rate of convergence is governed by
the eigenvalue with smallest modulus, which is now
\begin{equation}
\mu_1=-2\gamma(1-\epsilon)\; .
\label{eq:mu1nonergodic}
\end{equation}
Note that, contrary to the ergodic regime, the eigenvalue is always a
real number, so no oscillations occur, at least in the linearised
approximation.

\subsection{Marginal case $\epsilon=1$}

Let us proceed by approaching the marginal case $\epsilon=1$ from the
non-ergodic side, i. e. from below.
It might be instructive to cast the equations (\ref{eq:fors3s4}) in terms
of the eigenmodes of the linearised approximation. Namely, we can
introduce the functions
\begin{equation}
\begin{split}
\xi(t)&=\frac{1}{(2\epsilon-1)\gamma}\left(s_3(t)-\gamma\,s_4(t)-\frac{1}{2\gamma}\right)
\\
\eta(t)&=\frac{1}{(2\epsilon-1)\gamma}\left(-s_3(t)
+2\epsilon\gamma\,s_4(t)+\frac{1-\epsilon}{\gamma}\right)
\end{split}
\label{eq:defxieta}
\end{equation}
and express the equations (\ref{eq:fors3s4}) in the form
\begin{equation}
\begin{split}
\dot{\xi}&=-2\gamma(1-\epsilon)\,\xi+2\epsilon\gamma^3(2\epsilon\,\xi+\eta)\,\xi
\\
\dot{\eta}&=-\gamma\,\eta-2\epsilon\gamma^3(2\epsilon\,\xi+\eta)\,\xi
\end{split}
\label{eq:forxieta}
\end{equation}
The function $\xi(t)$ has a straightforward physical
interpretation: it describes the time evolution of the switching rate
of the external force. Indeed, inserting (\ref{eq:defxieta}) into
(\ref{eq:lambdathroughsses}) we get
\begin{equation}
\lambda(t)=2\gamma^3(1-2\epsilon)\,\xi(t)\; .
\end{equation}

The equations (\ref{eq:forxieta}) are a convenient starting point for
the investigation of the marginal regime. Taking $\epsilon=1$, the
linear term in the equation for $\dot{\xi}$ vanishes, while the linear
term in the equation for $\dot{\eta}$ remains. This suggests
that in the long-time regime the value of $\eta$ will be negligible
compared to $\xi$. This consideration will yield the leading term in the 
relaxation.

Thus, supposing $|\eta|\ll|\xi|$ we get the following approximate
equation for $\xi$ 
\begin{equation}
\dot{\xi}=4\gamma^3\,\xi^2
\end{equation}
which leads to the following asymptotic behaviour
\begin{equation}
\xi(t)\simeq -\frac{1}{4\gamma^3}\,\frac{1}{t}\quad ,\; t\to\infty
\quad .
\end{equation}
Now we must check the assumption that $\eta$ is negligible compared to
$\xi$. However, from (\ref{eq:forxieta}) we can see that the leading
term in the relaxation of $\eta$ is
\begin{equation}
\eta(t)\simeq\frac{1}{\gamma}\,\frac{1}{t^2}\quad ,\;  t\to\infty
\end{equation}
and the assumption is therefore consistent.

\begin{figure}[t]
\includegraphics[scale=0.8]{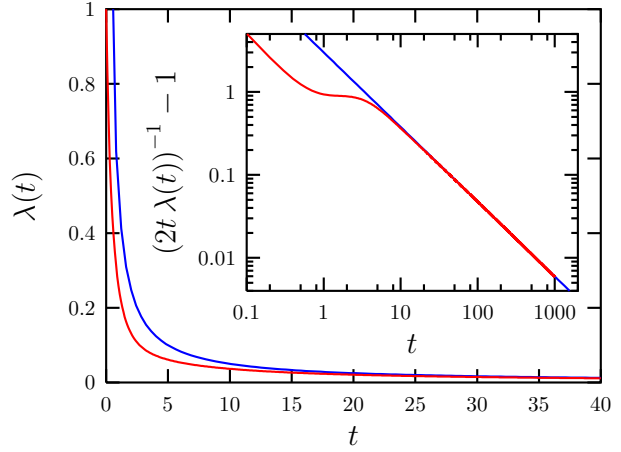}
\caption{%
Time evolution of the switching rate in the marginal regime
$\epsilon=1$, for  $\gamma=1$, $q=1$. The \slaninalineone\ is
the numerical 
solution, \slaninalinetwo\ %
the asymptotic analytical solution (\ref{eq:lambdamarginal}). In the
inset the deviation from the expression (\ref{eq:lambdamarginal}) is
shown (\slaninalineone). The \slaninalinetwo\ is
power dependence $3\times\, t^{-0.9}$.} 
\label{fig:lambda-critical}
\end{figure}

The consequence to draw is that in the marginal regime the relaxation
becomes power-law with exponent $-1$. Especially, the relaxation of
the switching rate follows the behaviour
\begin{equation}
\lambda(t)\simeq \frac{1}{2t}\quad,\; t\to\infty\; .
\label{eq:lambdamarginal}
\end{equation}
In Fig. \ref{fig:lambda-critical} we can compare  the numerical solution
with the asymptotic behaviour (\ref{eq:lambdamarginal}). 
We can see not only that that the function $\lambda(t)$ approaches
zero according to the power decay (\ref{eq:lambdamarginal}), but also
the corrections to the asymptotic behaviour can be well approximated by
a power. Indeed, from the inset in Fig. \ref{fig:lambda-critical} we
can see that 
\begin{equation}
\frac{1}{2t\,\lambda(t)}-1\simeq 3\,t^{-0.9}\quad,\; t\to\infty\; .
\label{eq:lambdamarginalcorrection}
\end{equation}
It is interesting to note that the power in the correction is not an
integer, so the naive expansion of the solution in powers of $t^{-1}$
cannot be used here. Instead, the behaviour
(\ref{eq:lambdamarginalcorrection})
suggests the expression in the form of a continued fraction
\begin{equation}
\lambda(t)= \cfrac{1}{a_1 t^{\alpha_1}
           +\cfrac{1}{a_2 t^{\alpha_2}
           +\cfrac{1}{a_3 t^{\alpha_3}
           +\dotsb
           }}}
\end{equation}
where the values $a_1=2$ and $\alpha_1=1$ are known exactly and the
next pair of parameters is estimated from the
numerical solution as $a_2\simeq 1/6$ and $\alpha_2\simeq 0.9$.

\section{Correlation functions}
\label{sec:correlationfunctions}

Additional information on the properties of the transition from
ergodic to non-ergodic behaviour which
occurs at the value $\epsilon=1$ can be gained from the two-time
correlation functions. Let us have $t>t_1>0$ and define the
correlation function
\begin{equation}
C(t,t_1)=\langle{\sf X}(t){\sf X}(t_1)\rangle\; .
\end{equation}
It can be expressed through the functions $s_1(t)$ to $s_4(t)$. The most
general formula is
\begin{equation}
\begin{split}
C(t,t_1)&=\langle{\sf X}_0\,\rangle{\rm e}^{-\gamma(t+t_1)}+\\
&+
\langle{\sf X}_0{\sf Q}_0\rangle\left[
{\rm e}^{-\gamma t_1}\,s_2(t) +{\rm e}^{-\gamma t}\,s_2(t_1)
\right]+\\
&+q^2\biggl[
2{\rm e}^{-\gamma(t-t_1)}\,s_4(t_1)+\\
&+\left(
\,s_2(t)-{\rm e}^{-\gamma (t-t_1)}\,s_2(t_1)
\right)
\frac{s_3(t_1)}{s_1(t_1)}
\biggr]
\end{split}
\end{equation}
although we suppose throughout this section that ${\sf X}_0=0$.

\begin{figure}[t]
\includegraphics[scale=0.8]{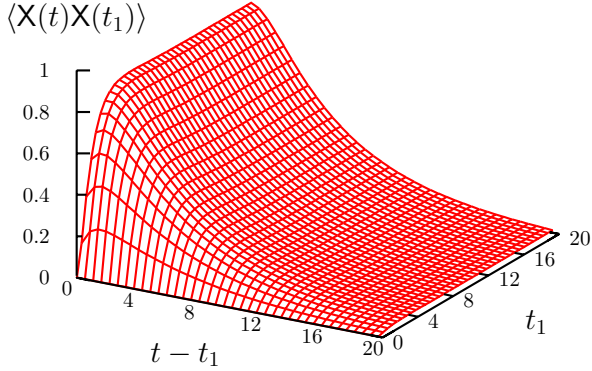}
\caption{Correlation function in the ergodic regime, for
$\epsilon=1.2$, $\gamma=1$, $q=1$.}
\label{fig:correl-ergodic}
\end{figure}

\begin{figure}[t]
\includegraphics[scale=0.8]{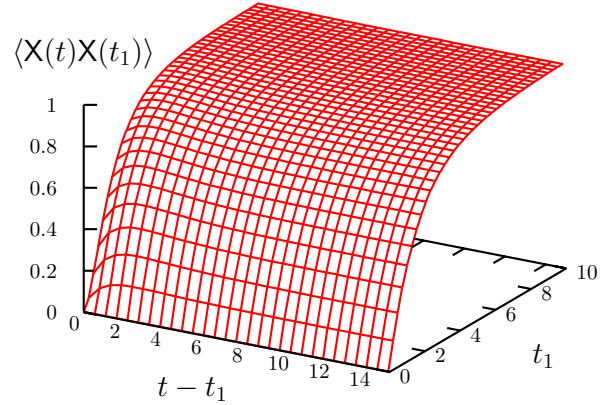}
\caption{Correlation function in the non-ergodic regime, for
$\epsilon=0.8$, $\gamma=1$, $q=1$.}
\label{fig:correl-nonergodic}
\end{figure}

We show in figures \ref{fig:correl-ergodic} (ergodic regime) and
\ref{fig:correl-nonergodic} (non-ergodic regime)
the evolution of correlation functions 
using the numerical solution for $s_1$ to $s_4$.
We can observe the damping of the correlations in the ergodic regime,
while in the non-ergodic regime the correlations converge to a finite
limit.
Let us now turn to the analytic investigation of the long-time
behaviour of the correlation function.

For long enough times we can suppose
that we are in the regime of exponential asymptotic relaxation of the
functions $s_1(t)$ to $s_4(t)$ and $\lambda(t)$, which is governed by
the eigenvalue closest to 0, as given by (\ref{eq:mu1ergodic}) and
(\ref{eq:mu1nonergodic}).

Let us start with the ergodic regime $\epsilon>1$. We find that for
both $t\to\infty$ and $t_1\to\infty$ the correlation function behaves
like
\begin{equation}
\begin{split}
C(t,t_1)\simeq \frac{q^2}{\gamma^2\epsilon(2-\epsilon)}
&\left[{\rm e}^{-\gamma(\epsilon-1)(t-t_1)}-\right.\\
&\left.-(\epsilon-1){\rm e}^{-\gamma(t-t_1)}\right]\; .
\end{split}
\end{equation}
Important feature of this result is that the correlation in asymptotic
regime depends only on the time difference $t-t_1$ and decays to zero
when this difference increases. This supports the picture of the
$\epsilon>1$ phase as a usual ergodic regime without long-time
correlations.

The situation is dramatically different for $\epsilon<1$. Technically
speaking, it is important that the function $\Lambda(t)$ has a finite
limit for large times,
$\lim_{t\to\infty}\Lambda(t)=\Lambda_\infty<\infty$.
For $\epsilon$ close to $1$ (i. e. $1-\epsilon\ll 1$)
the approach to this limit value can be written in the form
\begin{equation}
\Lambda(t)\simeq\Lambda_\infty-\theta\,{\rm e}^{-2\gamma(1-\epsilon)\,t}
\end{equation}
 where
$\theta$ is a constant depending on $\epsilon$.

For large times the correction will be small and we can formally write
the correlation function as expansion in powers of $\theta$. However,
we should bear in mind that it is not $\theta$ itself, which is small,
but the factor ${\rm e}^{-2\gamma(1-\epsilon)\,t}$ which appears
always together with $\theta$.

Finally, 
\begin{equation}
\begin{split}
C(t,t_1)&\simeq \frac{q^2}{\gamma^2}+
\frac{\theta q^2}{(\gamma-\mu)\gamma}\left(
{\rm e}^{-\mu\,t}
-{\rm e}^{-\mu\,t_1}
\right)-\\
&-\frac{2\theta q^2\mu}{(\gamma-\mu)(2\gamma-\mu)\gamma}
{\rm e}^{-\gamma(t-t_1)-\mu\,t_1}
+O(\theta^2)
\end{split}
\end{equation}
using $\mu=2\gamma(1-\epsilon)$ for shorter notation.

We can see that the qualitative difference from the ergodic regime
consists in the fact that for $\epsilon<1$ the correlation function
converges to a positive $\epsilon$-independent constant $q^2/\gamma^2$.
If we define the Edwards-Anderson order parameter
\begin{equation}
q_{\rm EA}=\lim_{\tau\to\infty}\lim_{t_1\to\infty}C(t_1+\tau,t_1)
\end{equation}
we can see that $q_{\rm EA}$ jumps discontinuously from the value
$q_{\rm EA}=0$ 
for $\epsilon>1$ to $q_{\rm EA}=q^2/\gamma^2$ for $\epsilon<1$. 
This observation represents another evidence that there is a
transition from ergodic regime to non-ergodic glassy regime at
$\epsilon=1$. As the Edwards-Anderson parameter is discontinuous at
the critical point, the transition should be classified as first-order
from this point of view. However, because we do not deal with an
equilibrium transition and the phenomenon is of purely dynamical origin,
the canonic classification of phase transition as first or second order
is of limited relevance here.  

\section{Critical behaviour at $\epsilon\to 1^-$}

We have already seen that it is possible to characterise the
glass transition at $\epsilon=1$ through the Edwards-Anderson order
parameter $q_{\rm EA}$. It has discontinuity at the transition, so the
corresponding critical exponent is 0.
Here we investigate another quantity, which can play the role of an
order parameter, being zero in the ergodic and non-zero 
in the non-ergodic phase.

\begin{figure}[t]
\includegraphics[scale=0.8]{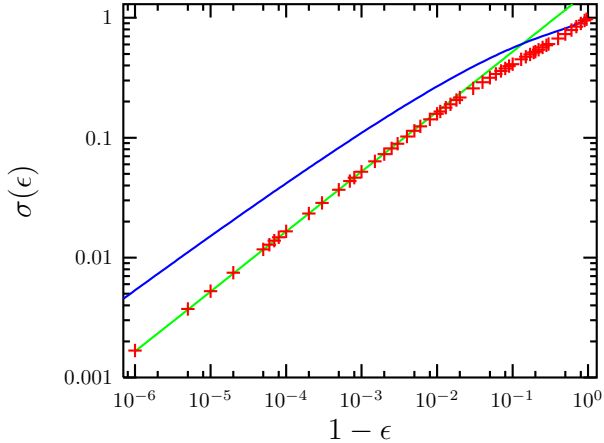}
\caption{%
The critical behavior at $\epsilon\to 1$. Points
(\textcolor{red}{$+$}) are results of numerical 
integration. The \slaninalineonethree\ %
is the approximation (\ref{eq:sigma-has-exp-half-analytically}), while
the \slaninalinetwo\ is the exact 
upper bound $\sigma_{\rm upper}(\epsilon)$ given by formula
(\ref{eq:sigmaupperbound}).  
}
\label{fig:sigma-vs-epsilon}
\end{figure}

The quantity in question will describe how the 
 initial conditions affect the
asymptotic vale of the average coordinate. We have already touched
 this point in  Sec. \ref{sec:nonergodic}.
Up to now we assumed that the initial condition for the noise is
such that $\langle{\sf Q}_0\rangle=0$. This assumption has the
consequence that both in 
ergodic and non-ergodic regime the average coordinate converges to
$0$. In this section we investigate the case $\langle{\sf Q}_0\rangle\ne0$.

From (\ref{eq:averageX}) and (\ref{seqn2})we
can see that 
\begin{equation}
\lim_{t\to\infty}\langle{\sf X}(t)\rangle=\frac{\langle{\sf
Q}_0\rangle}{\gamma}\,\sigma(\epsilon)
\end{equation}
 where we defined, as in Sec. \ref{sec:nonergodic},
\begin{equation}
\sigma(\epsilon)=\lim_{t\to\infty}s_1(t)
\label{eq:sigmaepsilon}
\end{equation}
stressing explicitly the dependence on $\epsilon$.
The equations (\ref{eq:fors3s4}) hold also in  case $\langle{\sf
Q}_0\rangle\ne0$. Therefore, we can proceed without 
further complications. We only need to plug the solution obtained the
same way as in the sections \ref{sec:glasstransition} and
\ref{sec:correlationfunctions} into the 
definition (\ref{sdef1}) of the function $s_1(t)$ and find
the limit (\ref{eq:sigmaepsilon}).

Let us first present the results of numerical integration. We will
turn to the analytic estimate afterwards.
The Fig. \ref{fig:sigma-vs-epsilon} shows the results of numerical
integration, indicating that asymptotically for $\epsilon\to 1$ the
behaviour follows the power law
\begin{equation}
\sigma(\epsilon)\sim (1-\epsilon)^\frac{1}{2}\; .
\label{eq:sigma-has-exp-half-empirically}
\end{equation}
The exponent $1/2$ is still observed only empirically and we do not
possess any proof that this is the exact value.

However, we may get some analytical argument in favour of this type of
behaviour 
from the equations (\ref{seqn3}), (\ref{seqn4}), and
(\ref{eq:lambdathroughsses}), which can be rewritten in slightly
different form. Defining new function $\psi(t)=s_3(t)-1/\gamma$ we can
write the set of equations for the pair $\psi(t)$ and $\lambda(t)$
\begin{equation}
\begin{split}
\dot{\psi}&=-\gamma\,\psi-\frac{1}{\gamma}\,\lambda-\lambda\psi
\\
\dot{\lambda}&=-2\gamma(1-\epsilon)\,\lambda+2\epsilon\gamma^2\,\lambda\psi
\end{split}
\label{eq:psilambda}
\end{equation}
with initial conditions $\psi(0)=-1/\gamma$, $\lambda(0)=\epsilon\gamma$.
If we further define $\Lambda_\infty=\int_0^\infty\lambda(t){\rm d}t$
and $\Psi_\infty=\int_0^\infty\psi(t){\rm d}t$, we can integrate both
LHS and RHS of the equations (\ref{eq:psilambda}) and obtain the
exact relation between $\Lambda_\infty$ and $\Psi_\infty$
\begin{equation}
\epsilon=-2\Lambda_\infty-2\epsilon\gamma^2\,\Psi_\infty\; .
\label{eq:LambdainfPsiinf}
\end{equation}
Knowing $\Lambda_\infty$ would solve the problem, because
$\sigma(\epsilon)={\rm e}^{-\Lambda_\infty}$. However,
in addition to (\ref{eq:LambdainfPsiinf}) we need some other 
condition. It can be established from the observation that the equation
for $\lambda(t)$ can be formally solved in the form
\begin{equation}
\lambda(t)=\epsilon\gamma\,\exp\left(-2\gamma(1-\epsilon)\,t
+2\epsilon\gamma^2\int_0^t\psi(t'){\rm d}t'\right)\; .
\end{equation}
and because $\psi(t)<0$, we have 
$\int_0^t\psi(t'){\rm d}t'>\Psi_\infty$.
Therefore, we can write the following upper bound
\begin{equation}
\sigma(\epsilon)<\exp\left(-\frac{1}{2}W_{\rm L}\left(
\frac{\epsilon\,{\rm e}^{-\epsilon}}{1-\epsilon}\right)\right)
\equiv\sigma_{\rm upper}(\epsilon) 
\label{eq:sigmaupperbound}
\end{equation}
where $W_{\rm L}(x)$ is the Lambert function defined by the equation
$W_{\rm L}(x)\,{\rm e}^{W_{\rm L}(x)}=x$.

The leading term in the asymptotic behaviour of the Lambert function
for large argument is $W_{\rm L}(x)\simeq\ln x$. If we use it as an
approximation for calculating the asymptotic behaviour of
$\sigma(\epsilon)$, starting with (\ref{eq:sigmaupperbound}) we
finally get
\begin{equation}
\sigma(\epsilon)\simeq\sqrt{\rm e}\,\sqrt{1-\epsilon}
\label{eq:sigma-has-exp-half-analytically}
\end{equation}
which is compatible with the behaviour
(\ref{eq:sigma-has-exp-half-empirically}). Actually, we can see in the Fig.
\ref{fig:sigma-vs-epsilon} that the approximation 
(\ref{eq:sigma-has-exp-half-analytically}) fits very well the results
from numerical integration and lies much closer than the exact upper
bound (\ref{eq:sigmaupperbound}). Thus, we conjecture that the formula
(\ref{eq:sigma-has-exp-half-analytically})  is  in fact the correct
asymptotic behaviour for $\epsilon\to 1$.

To sum up, in the regime $\epsilon>1$, the initial conditions are irrelevant 
for long-time dynamics, as expected in the ergodic phase. On the other
hand, for $\epsilon<1$, we observe that the asymptotic value of the average
coordinate depends on the initial condition for the noise, which is
yet another signature of ergodicity breaking. The
factor $\sigma(\epsilon)$ measures the sensitivity to initial
conditions: it vanishes in ergodic phase but remains
non-zero in non-ergodic phase. So, it may be considered as a kind of
order parameter. Close to
the critical point $\epsilon=1$ it approaches zero continuously as a
power with critical exponent $1/2$.

\begin{figure}[t]
\includegraphics[scale=0.8]{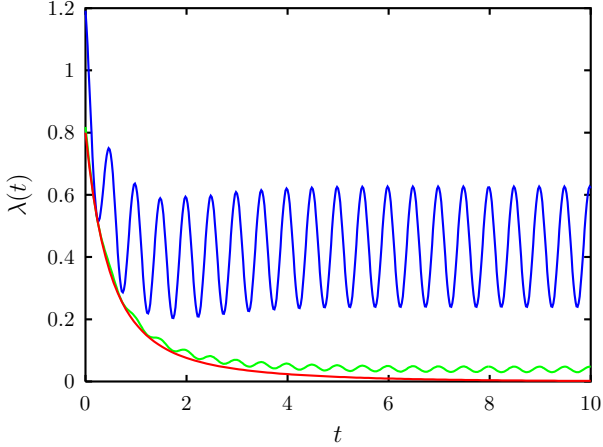}
\caption{%
Response of the switching rate to harmonic external 
driving force
for $\gamma=1$, $q=1$, $\epsilon=0.8$, $\omega=2\pi$ and $F_0=0.7$
(\slaninalinetwo) and $F_0=0.15$ (\slaninalinethreefour). 
The \slaninalineone\  %
shows the time dependence in absence of
the external driving force.}
\label{fig:responselambda}
\end{figure}

\section{Response to harmonic perturbation}

Let us investigate now the response of the particle to the external
driving force.  Adding the additional term $F(t)=F_{0}\cos(\omega t)$
at the right 
hand side of Eq. (\ref{eq:langevin}),  one can repeat, {\it mutatis
mutandis\/}, all steps leading to the system of equations for the
functions $s_1(t)$ to $s_4(t)$.  We find that the equations
(\ref{seqn1})-(\ref{seqn4})
hold unchanged, while the influence of the external force modifies the
expression (\ref{eq:lambdathroughsses}) for $\lambda(t)$.
Actually, in the present case one gets
\begin{equation}
\begin{split}
\frac{\lambda(t)}{\epsilon\gamma}&=
1-2\gamma\,s_3(t)+2\gamma^2\,s_4(t)+\\
&+\frac{1}{q^2}\left(
\gamma\,{\rm e}^{-\gamma t}\int_0^t F(t')\,{\rm e}^{\gamma t'}\,{\rm d}t'
-F(t)
\right)^2\quad .
\end{split}
\label{eq:lambdathroughssesforce}
\end{equation}
We assumed $\langle{\sf Q}_0\rangle=0$ and $\langle{\sf X}_0\rangle=0$ here.

First quantity to study is the response of the average coordinate. We
find
\begin{equation}
\begin{split}
\langle{\sf X}(t)\rangle=&
\langle{\sf Q}_0\rangle\,
{\rm e}^{-\gamma t}\int_0^t{\rm e}^{-\Lambda(t')+\gamma t'}\,{\rm d}t'
+\\
&+
{\rm e}^{-\gamma t}\int_0^t F(t')\,{\rm e}^{\gamma t'}\,{\rm d}t'
\quad .
\end{split}
\end{equation}
Obviously enough, in the stationary regime the average coordinate oscillates
around $0$ with the same frequency as the driving force $F(t)$. We
arrive at the standard Debye-type dynamic susceptibility
\begin{equation}
\chi(t-t')=\Theta(t-t'){\rm e}^{-\gamma(t-t')}\quad .
\end{equation}
Thus, the exact response in terms of the coordinate is linear.
This behaviour also does not depend on the value of $\epsilon$.

\begin{figure}[t]
\includegraphics[scale=0.8]{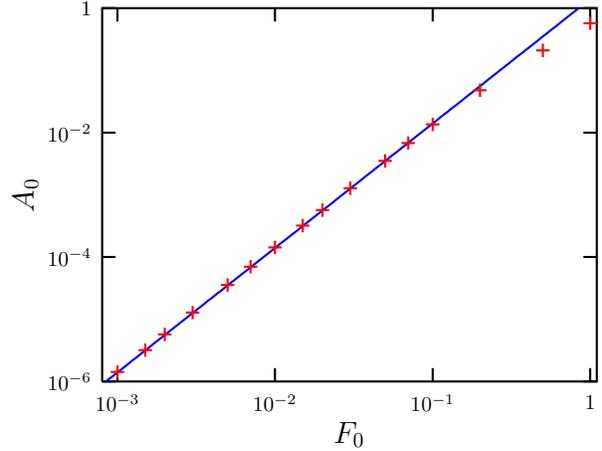}
\caption{%
Dependence of the constant term in the Fourier series
(\ref{eq:lambdafourierseries}) for $\lambda_{\rm st}(t)$ on the
amplitude of the driving force ($+$), in the regime $\epsilon<1$. 
The parameters are $\gamma=1$,
$q=1$, $\epsilon=0.8$ and the frequency is $\omega=\pi/2$.   
Full line is the function $1.4\,{F_0}^2$.}
\label{fig:averlambda-vs-fzero}
\end{figure}

The situation becomes much more complicated when we turn to
quantities, which depend non-linearly on the coordinate, especially
the switching rate $\lambda(t)$. 
We solved numerically the set of equations (\ref{seqn1}) to (\ref{seqn4})
and (\ref{eq:lambdathroughssesforce}). 
We can see in Fig. \ref{fig:responselambda} the evolution of the
function $\lambda(t)$ within the non-ergodic regime, with
$\epsilon=0.8$. Comparing the behaviour with the evolution in absence
of the  external harmonic perturbation we can see that the switching
rate $\lambda(t)$ does not approach zero any more, but it oscillates around
some finite value, which we will denote $A_0$. This qualitative
feature holds for whatever small 
external force. However, when the amplitude $F_0$ of the external
field goes to zero, also the value of $A_0$ approaches zero according
to $A_0\sim {F_0}^2$, as can be seen from
Fig. \ref{fig:averlambda-vs-fzero}.  
Generalising the linear stability analysis of
Sec. \ref{sec:fixedpoits} to harmonic oscillations, we conclude that
the external field continuously shifts the fixed point with
$\lambda=0$, (i. e. also $A_0=0$) to a position with positive $A_0$,
but the value of the shift vanishes when the amplitude of the
perturbation goes to zero. 

\begin{figure}[t]
\includegraphics[scale=0.8]{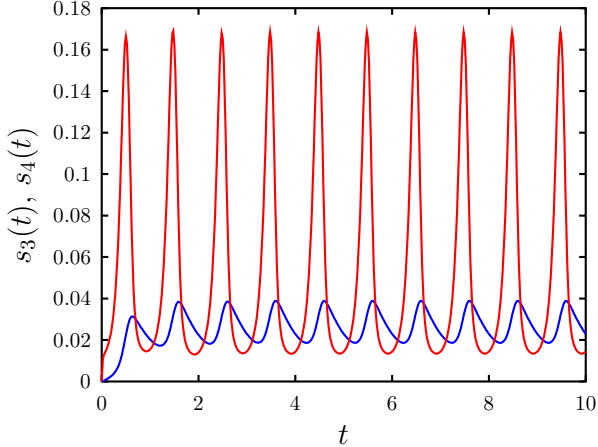}
\caption{%
Response of the functions $s_3(t)$ (\slaninalineone) and $s_4(t)$
(\slaninalinetwo) to harmonic 
external  
driving force
for  $\gamma=1$, $q=1$, $\epsilon=0.8$, $\omega=\pi$ and $F_0=10$.}
\label{fig:responses3s4}
\end{figure}

 Figure \ref{fig:responses3s4} exemplifies  the evolution of the
functions $s_3(t)$ and $s_4(t)$. We can clearly see that the
oscillations are not harmonic. Generally, 
in the stationary regime these functions are non-harmonic but periodic
with the doubled frequency $2\omega$.
Thus, the same holds also for the function $\lambda(t)$.
Using (\ref{eq:lambdathroughssesforce}) the stationary response in
terms of the switching rate can be written as Fourier series
\begin{equation}
\lambda_{\rm st}(t)=A_0+\sum_{k=1}^\infty\left(
 A_k\sin 2k\omega t+
 B_k\cos 2k\omega t\right)\quad .
\label{eq:lambdafourierseries}
\end{equation}
The amplitudes of the harmonic
components $A_0$, $A_k,B_k$, 
$k=1,2,...$, satisfy a complicated infinite set of quadratic equations.

\begin{figure}[t]
\includegraphics[scale=0.8]{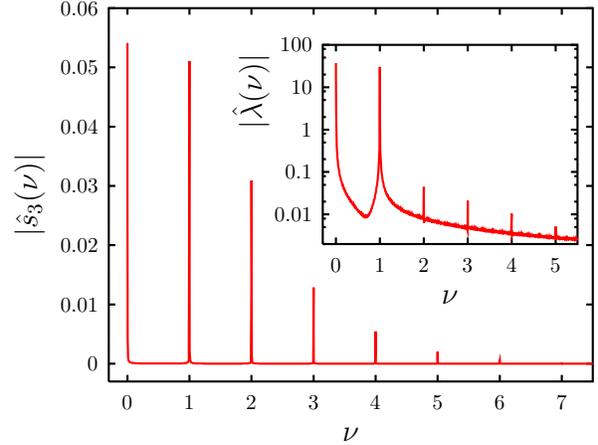}
\caption{%
Fourier transform of the stationary oscillations of the function
$s_3(t)$ under the influence of harmonic external 
driving force
for $\epsilon=0.8$, $\omega=\pi$ and $F_0=10$. It was calculated using
the fast Fourier transform algorithm. In the inset
the Fourier transform of oscillations of the 
switching rate. The finite width of the peaks, the noise and the
continous part of the spectrum in 
the inset are due to numerical imprecision of the fast Fourier
transform procedure.}
\label{fig:responsespectrum}
\end{figure}

To asses the weight of the higher harmonics we performed the fast
Fourier transform of the time evolutions obtained by numerical 
solution, throwing away the initial transient regime. To illustrate
the presence of higher harmonics we chose the function $s_3(t)$. The
modulus of its Fourier transform $\hat{s}_3(\nu)=\int s_3(t)\,{\rm
e}^{-2\pi{\rm i}\,\nu t}\;{\rm d}t$ is shown in
Fig. \ref{fig:responsespectrum}. We can clearly see the peaks 
at the multiples of the basic frequency. We can also observe that the
higher harmonics have quite considerable weight. In the inset of
Fig. \ref{fig:responsespectrum} we show also the Fourier transform of
the function $\lambda(t)$. Here, the higher harmonics are much less
pronounced. 

The most important feature of the time evolution under the influence
of external harmonic force is the observation
that the switching rate remains always positive. This leads to the
already mentioned fact that whatever is the coupling strength
parameter $\epsilon$, the mean coordinate in stationary regime  
oscillates around zero, irrespectively of the initial
conditions. However, this is the signature of ergodicity, so the
glassy behaviour disappears under the influence of arbitrarily small
external perturbation. Such a behaviour was already observed also in the
model of sheared colloid \cite{cates_02}.

Both observations
can be easily understood. The external force drags the particle back and
forth.  Once moving, the particle induces through the back-reaction
(\ref{eq:lambda}) the 
fluctuations of the environmental  force, which prevents the system
from freezing in a non-ergodic state. This holds for any positive amplitude
of the driving force. However, when the force diminishes, there is
still longer transient period, where the system apparently 
relaxes toward the arrested state, as can be seen qualitatively in
Fig. \ref{fig:responselambda}. 
In the limit of infinitesimally small driving, the
transient time blows up and the asymptotic state
corresponds to the dynamically arrested
state. This picture is consistent with the view of glassiness as
a purely dynamical phenomenon.

\begin{figure}[t]
\includegraphics[scale=0.8]{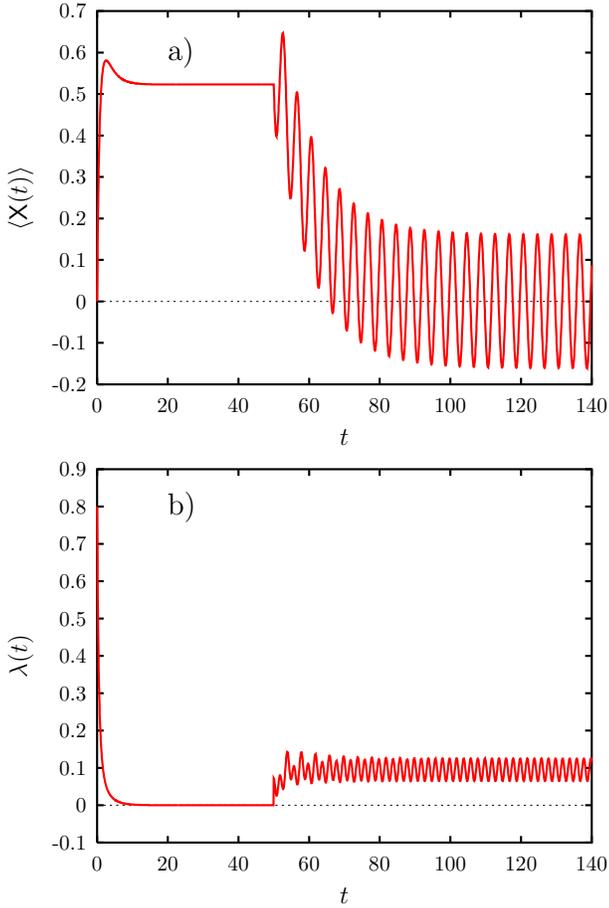}
\caption{%
Time evolution of the verage coordinate (a) and switching rate (b) in
the regime 
$\epsilon<1$, when the external harmonic perturbation was switched on at
time $t=50$. The parameters are $\gamma=1$,
$q=1$, $\epsilon=0.8$, the amplitude of the external perturbation is
$F_0=0.3$  and the frequency is $\omega=\pi/2$. The initial condition
is $\langle{\sf Q}_0\rangle=1$.
}
\label{fig:response-later-0.8}
\end{figure}

We also observed the response of the system to a signal, which is
switched on only after the system relaxed very close to the arrested
state. The results can be seen in Fig. \ref{fig:response-later-0.8}.
Initially, the switching rate relaxes toward zero, but after the
perturbation it settles on oscillating behaviour. The average 
coordinate initially approaches non-zero value (we have chosen initial
condition $\langle{\sf Q}_0\rangle>0$), but the perturbation brings it
to oscillations around zero. This can be interpreted as a schematic
picture of shear thinning, although the model is too much simplified to
account for the shear thinning quantitatively.

\section{Conclusions}

Motivated by the dynamical arrest phenomenon in colloids
we formulated and solved a stochastic dynamical model, where the coordinate of
a single particle
 evolves under the influence of stochastic
environmental force. The back-reaction couples the switching rate of
the force to the average of the square of the velocity 
of the particle. The strength of the coupling $\epsilon$ is the crucial
parameter which determines the behaviour of the system. The problem
reduces to a set of coupled non-linear 
differential equations, which was investigated both analytically and
numerically.

The back-reaction induces a phase transition from the
ergodic phase for $\epsilon>1$ to the non-ergodic glassy state for
$\epsilon<1$. The transition is observed qualitatively in the
behaviour of the switching rate, decaying to zero in non-ergodic
state, while staying positive in the ergodic state.  The
Edwards-Anderson parameter, established from the two-time correlation
functions, is discontinuous at the transition; it is zero in ergodic
phase ($\epsilon>1$), while for $\epsilon<1$ it acquires
finite value independent of $\epsilon$. The critical point
$\epsilon=1$ is characterised by power-law decay of the switching
rate. The leading term $\sim t^{-1}$ in the long-time behaviour was
calculated  analytically. 

We investigated the critical behaviour at $\epsilon\to 1$ through the
dependence of the average coordinate in the long-time limit
on the initial condition. We find that the asymptotic value of the
average coordinate is proportional to the average initial value of the
force, where the proportionality factor $\sigma(\epsilon)$ is singular
at the transition. In the ergodic phase we have $\sigma(\epsilon)=0$
identically, while in the non-ergodic phase, close to the transition,
we found $\sigma(\epsilon)\sim(1-\epsilon)^{1/2}$ for 
$\epsilon\to 1^-$.

Therefore, we find a situation quite unusual from the point of view of
static equilibrium phase transition. Indeed, we have two variables,
which may be 
considered as order parameter, namely the Edwards-Anderson parameter
and the quantity $\sigma(\epsilon)$. While the former is discontinuous
at the critical point, thus indicating first-order transition, the
latter is continuous, suggesting second-order transition. the
discrepancy is to be attributed to purely dynamical nature of the 
transition.

Finally, we investigated the response of the system to harmonic
external perturbation. We found that the exact response of the coordinate is
linear. On the other hand, the
response 
of the variables which are quadratic functions of the coordinate and
velocity, like the switching rate, was non-linear and generically
contains all higher harmonics, as was seen in the Fourier transform of
the signal. We also observed that arbitrarily weak external
perturbation is sufficient to ``melt'' the non-ergodic glassy state
and bring it back to ergodic behaviour. We may relate this feature to
the notion of stochastic stability \cite{parisi_00}; in this view our
system is not 
stochastically stable. However, our finding is in accord with
previously observed behaviour
of sheared colloids \cite{cates_02}. It makes also connection to the
rheological properties of thixotropic fluids
\cite{vol_nit_hey_reh_02,cou_ngu_huy_bon_02}, although our model is
too simplified to give quantitative predictions in this direction. 

The back-reaction mechanism described by (\ref{eq:lambda}) represent
the simplest choice. One may ask what would happen if we tried another
prescription. We expect that the methods used here will be as well
applicable if we generalise (\ref{eq:lambda}) as $\lambda(t)=\langle
F({\sf V}^2(t))\rangle$ for an analytic function $F(x)$. More
complicated situation would appear if the dependence was non-local in
time, e. g. of the form $\lambda(t)=\int^t\langle
{\sf V}(t){\sf V}(t'))\rangle\,K(t-t')\,{\rm d}t'$ with some
kernel $K(t)$.  Such an approach would bring our model closer to the
well-studied Mode Coupling equations, but it goes beyond the scope of
the present work.

\begin{acknowledgement}
We want to dedicate this paper to the memory of Prof. Vladislav
\v{C}\'apek, a passionate theoretical physicist, our dedicated teacher
and good friend, who passed away shortly 
before   this work was completed.

 This work was supported by the project No. 202/00/1187
of the Grant Agency  of the Czech Republic. 
\end{acknowledgement}
{}
\end{document}